\documentclass[10pt,conference]{IEEEtran}
\IEEEoverridecommandlockouts
\usepackage{cite}
\usepackage{amsmath,amssymb,amsfonts}
\usepackage{algorithmic}
\usepackage{graphicx}
\usepackage{textcomp}
\usepackage{xcolor}
\usepackage{url}
\usepackage{comment}
\usepackage{soul}
\usepackage{booktabs}
\usepackage{multirow}

\usepackage{lscape}

\usepackage{tikz}
\usepackage{textcomp}
\usepackage{hyperref}
\usepackage{lipsum}

\newcommand\copyrighttext{%
  \footnotesize \textcopyright~  2023 IEEE.  Personal use of this material is permitted.  Permission from IEEE must be obtained for all other uses, in any current or future media, including reprinting/republishing this material for advertising or promotional purposes, creating new collective works, for resale or redistribution to servers or lists, or reuse of any copyrighted component of this work in other works. The official version can be found at
  \href{https://dx.doi.org/10.1109/DCOSS-IoT58021.2023.00081}{https://dx.doi.org/10.1109/DCOSS-IoT58021.2023.00081}}
\newcommand\copyrightnotice{%
\begin{tikzpicture}[remember picture,overlay]
\node[anchor=south,yshift=10pt] at (current page.south) {\fbox{\parbox{\dimexpr\textwidth-\fboxsep-\fboxrule\relax}{\copyrighttext}}};
\end{tikzpicture}%
}

\usepackage{pifont} 
\newcommand{\xmark}{\textcolor{red}{\ding{54}}}
\newcommand{\cmark}{\textcolor{green}{\ding{52}}}
\newcommand{\dmark}{\textcolor{blue}{\ding{110}}}

\sethlcolor{white}

\def\BibTeX{{\rm B\kern-.05em{\sc i\kern-.025em b}\kern-.08em
    T\kern-.1667em\lower.7ex\hbox{E}\kern-.125emX}}

\IEEEoverridecommandlockouts
\IEEEpubid{\makebox[\columnwidth]{XXX-X-XXXX-XXXX-X/22/\$31.00~\copyright{}2023 IEEE \hfill} \hspace{\columnsep}\makebox[\columnwidth]{ }}

\begin{document}

\title{
Towards Cyber Threat Intelligence for the IoT
}

\author{\IEEEauthorblockN{Alfonso Iacovazzi, Han Wang, Ismail Butun, Shahid Raza}
\IEEEauthorblockA{
RISE Research Institutes of Sweden,
  Isafjordsgatan 22, 164 40
 Kista, Sweden \\
 }
}

\maketitle
\copyrightnotice


\begin{abstract}

With the proliferation of digitization and its usage in critical sectors, it is necessary to include information about the occurrence and assessment of cyber threats in an organization’s threat mitigation strategy. This Cyber Threat Intelligence (CTI) is becoming increasingly important, or rather necessary, for critical national and industrial infrastructures. Current CTI solutions are rather federated and unsuitable for sharing threat information from low-power IoT devices. This paper presents a taxonomy and analysis of the CTI frameworks and CTI exchange platforms available today. It proposes a new CTI architecture relying on the MISP  Threat Intelligence Sharing Platform customized and focusing on IoT environment. The paper also introduces a tailored version of STIX (which we call tinySTIX), one of the most prominent standards adopted for CTI data modeling, optimized for low-power IoT devices using the new lightweight encoding and cryptography solutions. The proposed CTI architecture will be very beneficial for securing IoT networks, especially the ones working in harsh and adversarial environments.

\end{abstract}



\section{Introduction}
According to National Institute of Standards and Technology (NIST) \cite{johnson2017cyber}, a cyber threat is identified as follows: ``Any circumstance or event with the potential to adversely impact organizational operations (including mission, functions, image, or reputation), organizational assets, individuals, other organizations, or the Nation through an information system via unauthorized access, destruction, disclosure, or modification of information, and/or denial of service.''

Cyber Threat Intelligence (CTI) is the data and/or useful information gathered by the Computer Emergency Response Team (CERT
) of each organization or entity. It consists of information such as vulnerabilities, Time To Penetrate/Compromise (TTP), threat actor, network/incident logs, exploited targets, incident indicators, etc. In the past, manual exchange of such CTI data among \hl{CERTs} has been quite in common up-until exchange of CTI data was realized by means of automated CTI data exchange platforms.

\hl{CTIs commonly represent Indicator of Compromise (IoC)} for formalizing and/or representing threat actors. As Zhao et al. \cite{zhao2020cyber} states, IoCs and CTIs are proactive measures that are \hl{handy} for CERTs: ``Different from the well-known security databases (e.g., CVE, ExploitDB), CTI can facilitate organizations to proactively release more comprehensive and valuable threat warnings (e.g., malicious IPs, malicious DNS, malware and attack patterns, etc.) when a system encounters suspicious outsider or insider threats.'' Thus, the collection, analysis, and dissemination of CTI are essential for effective cyber defense. However, many organizations struggle to create CTI on their own, and even those that do often lack the resources to keep up with the rapidly evolving threat landscape.


To address these challenges, CTI exchange and sharing platforms have emerged as a solution. These platforms allow organizations to share CTI with each other, collaborate on analysis, and benefit from the collective knowledge and expertise of a community of users. CTI exchange and sharing platforms can be public or private, with some focused on specific industries or regions, while others are more open and inclusive.
One of the primary benefits of CTI exchange and sharing platforms is that they enable organizations to access and contribute to a wealth of CTI that they would not have been able to generate on their own. By pooling resources and expertise, organizations can quickly and effectively respond to emerging threats and prevent cyberattacks. 
One example of a CTI exchange and sharing platform is MISP (Malware Information Sharing Platform)~\cite{wagner2016misp} which is open source and widely used by research organizations and private companies to exchange threat intelligence.

CTI exchange and sharing platforms require (i) data models to represent IoCs and (ii) protocols exchanging IoCs between different systems. STIX (Structured Threat Information eXpression)  and TAXII (Trusted Automated eXchange of Indicator Information) are two prominent standards, defined by the standards body ``OASIS Open,''\footnote{\url{https://www.oasis-open.org/}} that are used in CTI exchange and sharing platforms. STIX is a language for representing cyber threat intelligence in a structured and machine-readable format~\cite{jordan2021stix}, while TAXII is a protocol for exchanging that information~\cite{jordan2021taxii}.

According to our pre-evaluations, the STIX and TAXII standards are too heavy (demanding in terms of processing power and memory, etc.) for the small IoT devices which mostly run on limited battery resources. As such, improvements on these proposed protocol standards are needed to make them lightweight and therefore suitable for the IoT devices.

Here, in this work, we propose a new CTI sharing architecture suitable for IoT environments. We introduce a lightweight version of the STIX data model which can be processed and handled by constrained devices. The proposed data model -- which we call ``tinySTIX'' -- together with the adequate IoC sharing models -- ``tinySTIX over COAP'' -- leverages on protocols developed for constrained devices: Concise Binary Object Representation (CBOR) for the CTI data representation, CBOR Object Signing and Encryption (COSE) for the security services, Constrained Application Protocol (CoAP) for the  transfer of CTI objects, and Object Security for Constrained RESTful Environments (OSCORE) for the end-to-end protection.
In addition, we present how the MISP sharing platform need to be customized to handle lightweight protocols in order to cater to the needs of the IoT-based CTI services.


The rest of the paper is as follows. Section \ref{Background} describes the architecture and related basics of CTI data format and CTI distribution platform. Whereas, Section~\ref{Taxonomy} provides a thorough analysis and taxonomy of the related work in the given field, especially on the CTI formats and distribution platforms in a comparative way. Section~\ref{CTIforIoT} proposes a new CTI platform designed for IoT environments and incorporating the tailored tiny version of STIX and the IoT-specific protocols. Finally, the overall work is concluded in Section~\ref{Conclusion}.

\section{Background} \label{Background}
This section covers the most prominent CTI format, STIX; and also the distribution framework associated with it, TAXII.
\subsection{Structured Threat Information Expression (STIX)}
STIX \cite{jordan2021stix} is an open source language and serialization format used to exchange CTI data among peers and groups, \hl{details of which are summarized as follows:} 

\subsubsection{Why is it important?}
STIX helps us easily contributing to and consuming from the CTI; as all dimensions of suspicion, compromise and attribution can be emphasized with descriptive relations along with the object identifiers. Moreover, it can be stored as a JSON file in a machine readable format and also can be visualized in a graphical representation which might ease the work of the analyzer. It is an open source software and allows seamless integration with other available tools and products on the market.

\subsubsection{STIX Variants}
CybOX (Cyber Observable eXpression) is a standardized language for describing and sharing cyber observables, which are pieces of information related to cyber threats, such as network traffic, file attributes, and registry keys. CybOX was developed by the Mitre Corporation and firstly released in 2010. However, the OASIS CTI technical committee\footnote{\url{https://www.oasis-open.org/committees/tc_home.php?wg_abbrev=cti}} later decided to merge its specifications into STIX and  CybOX objects are now referred to as STIX Cyber Observables. 
There are several STIX variants available. For instance, STIX 2.x represents the latest version and requires implementations to support JSON serialization. On the other hand, STIX 1.x was defined using XML. Although both XML and JSON have benefits, the CTI OASIS technical committee determined that JSON was more lightweight, and sufficient to express the semantics of CTI information. STIX 2.x (based on JSON) is simpler to use and increasingly preferred by developers nowadays.





\subsubsection{STIX Domain Objects (SDOs)}
STIX allows categorization of useful information with special attributes to be populated as Objects. SDOs refer to chaining several objects via relationships to allow CTI representations. The structured nature of the STIX architecture allows it to define relationship between constructs. For example the TTP  can be related to a specific threat actor. Explanations of most relevant SDOs are as follows:



\begin{itemize}
\item \textbf{Campaign}: These are the instances of the threat actors.
\item \textbf{CourseOfAction}: Recommendation from a producer of CTI to a consumer on the actions that they might take in response to that specific intelligence.
\item \textbf{ExploitTarget}: This type defines the vulnerabilities in software, systems, networks, etc. that are targeted for exploitation by the TTP of a ThreatActor.
\item \textbf{Incident}: Represents a single STIX Incident.
\item \textbf{Indicator}: A pattern to be used in detecting malicious cyber activity.
\item \textbf{Observable}: Also referred to as STIX Cyber-observable Objects (SCOs). Holds and helps distribution of information about cyber security related entities such as files, systems, and networks.
\item \textbf{ThreatActor}: These are characterizations of adversaries that possess threat abilities due to their past actions.
\item \textbf{TTP}: TTP (Tactics, Techniques, and Procedures) is a representation of the behavior of cyber attackers.
\end{itemize}

Other available SDOs are: Attack pattern, grouping, identity, infrastructure, intrusion set, location, malware, malware analysis, note, opinion, report, tool, and vulnerability.

\subsubsection{STIX Relationship  Objects (SROs)}
STIX allows linking SCOs and STOs through SROs. There are two categories defined:

\begin{itemize}
\item \textbf{Relationship:} They are leveraged to link together 2 SDOs or SCOs in order to describe how they are related to (connection types) each other.
\item \textbf{Sighting:} Specifies that something in CTI has observed.
\end{itemize}



\subsubsection{STIX backwards compatibility and interoperability}
STIX imports and leverages the following constituent schemas:
\begin{itemize}
\item Cyber Observable eXpression (CybOX™) v1.0
\item Indicator Exchange eXpression (IndEX™) v0.4
\item Common Attack Pattern Enumeration and Classification (CAPEC™) v2.5
\item Malware Attribute Enumeration and Characterization (MAEC™) v2.1
\item Incident Object Description and Exchange Format (IODEF) v1.0
\item Data\_Marking v0.3
\end{itemize}

\subsubsection{STIX's features}
Some prominent features of the STIX are as follows~\cite{jordan2021stix}:
\paragraph{Flexibility}
``STIX adheres to a policy of allowing users to employ any portions of the standardized representation that are relevant for a given context and avoids mandatory features wherever possible.''

\paragraph{Extensibility}
``The STIX design intentionally builds in extension mechanisms for domain specific use, for localized use, for user-driven refinement and evolution, and for ease of centralized
refinement and evolution''

\paragraph{Automate-ability}
``The STIX design approach intentionally seeks to maximize structure and consistency to support machine-process-able automation.''

\paragraph{Readability}
``The human readability is necessary for clarity and comprehensibility during the early stages of development and adoption, and for sustained use in diverse environments going forward.''

\subsection{TAXII}
TAXII is a framework which provides information on how to share the cyber-threat data. The sharing is executed by using HTTPS via message and exchange. There are 3 TAXII data sharing models \cite{jordan2021taxii}:
\begin{enumerate}
\item \textbf{Peer to Peer:} In this model, two or more entities communicate and exchange the information directly.
\item \textbf{Hub and Spoke:} One source is selected as the primary data source, whereas the entities communicate and exchange information with this specific data hub (repository).
\item \textbf{Publish and Subscribe:} Apart from the Hub and Spoke model, in this model the repository publishes the data whereas the subscribers consume it.
\end{enumerate}

TAXII defines a RESTful API and a set of requirements for TAXII Clients and Servers. There are 2 primary services in TAXII to support the data sharing models mentioned above:
\begin{enumerate}
\item \textbf{Collection:} Describes an interface to a logical repository of CTI objects, provided by the TAXII Server; the clients and servers of the TAXII exchange their CTI data in a challenge-response model. A collection allows a CTI producer to host a set of CTI data that can be requested by the CTI consumers.
\item \textbf{Channel:} They are maintained by a TAXII Server. Clients exchange information with other TAXII clients in a publish-subscribe model via channels. A channel allows producers to push data to many consumers as well as consumers to receive data from many producers.
\end{enumerate}

\section{Taxonomy and Analysis of the CTI Technologies} \label{Taxonomy}
In this section, the readers are provided with comprehensive taxonomy on CTIs and the technologies used to process them.

\subsection{Cyber Threat Intelligence (CTI) Frameworks}\label{frameworks}
We investigate CTI frameworks under three categories: ``data source'', ``way of distribution'', and ``formats'': 

\subsubsection{Data source}
According to the source of data, CTIs can be categorized into 3 classes~\cite{farnham2013tools}:
\begin{itemize}
\item \textbf{Internal}: This threat category includes CTIs that are collected from within the same organization. An example would be security reports from Security Incidence and Event Management (SIEM) systems, or more importantly from computer forensic analysis, which can reveal intelligence that is not visible otherwise and might be crucial while detecting other intrusions.
\item \textbf{Community}: This threat category includes any CTI shared via a trusted relationship with multiple members with a shared interest, such as special-interest groups consisting of organizations from the same industry sector.
\item \textbf{External}: This threat category encompasses CTI from sources outside an organization that are excluding the community members. External category is further grouped into 2 classes: Private and public. Private CTI sources operate on paid subscription model, whereas public CTI sources are publicly available to everyone.
\end{itemize}

\subsubsection{Way of distribution}
According to the way of distributing the data, CTIs can be categorized into 4 classes~\cite{Celerium2023CTI}:
\begin{itemize}
\item \textbf{Author/Publisher}: This is the actual source of the CTI data, which is emanating from cyber analysts towards the CTI subscribers.
\item \textbf{Subscriber}: Physical entities or individuals that are subscribed to various publishers and collect CTI from them.
\item \textbf{Peer-to-Peer}: This is an automated (M2M) way of publish/subscribe system. For instance connecting an enhanced firewall to CTI publishers via STIX/TAXII.
\item \textbf{Hub \& Spoke}: Information Sharing and Analysis Centres (ISACs) and  Information Sharing and Analysis Organizations (ISAOs) collect, optionally anonymize, and re-distribute the CTIs to member organizations.
\end{itemize}



\begin{table*}[htbp]
\footnotesize
\caption{Comparison of CTI Data Formats}
\begin{center}
\begin{tabular}{ccccc}
\toprule
\textbf{CTI data} & \textbf{Vendor/} & \textbf{Data format} & \textbf{Highlights for / summary of} & \textbf{IoT}\\ 
\textbf{format name} & \textbf{owner}  & \textbf{and/or protocol} & \textbf{each format} & \textbf{implementability}\\
\midrule
IODEF~\cite{danyliw2007incident} &  MILE & XML & object-oriented structured format, & NO\\
  &  &   & composed of 47 classes (unique features) & \\
\midrule
IODEF-SCI~\cite{takahashi2014incident} &  MILE & XML & extension to the IODEF standard that & NO\\
&  &  & adds support for additional data & \\
\midrule
RID~\cite{moriarty2012real} &  MILE & XML (XML1.0) & RID is defined in RFC 6545 and the transport of & NO\\
&   &  & RID messages over HTTP/TLS is defined in RFC 6546 & \\
\midrule
CyBOX~\cite{barnum2012cybox} &  MITRE & XML & 70 defined objects are defined, & NO \\
&  &  & supports X509 Certificate & \\
\midrule
STIX 1.x~\cite{jordan2021stix} &  MITRE & XML & indicator patterns are expressed using XML syntax. & NO \\
&  &  &  even simplest patterns are difficult to create and to understand & \\
\midrule
STIX 2.x~\cite{jordan2021stix} &  MITRE & JSON & STIX 2.x requires JSON serialization to be supported by  & YES\\
&  &  &  implementations, if written in the STIX patterning language, & \\
&  &  &  would be more compact and easier to read. &\\
\midrule
TAXII ~\cite{jordan2021taxii} &  MITRE & HTTPS & supports multiple sharing models, including variations of  & NO\\
&  &  & ‘hub and spoke’ as well as ‘peer to peer’. & \\
\midrule
OTX~\cite{OTX2023} & open & Pulses & The world's largest open CTI community that enables & INCONCLUSIVE\\
& community &  &  collaborative defense with actionable, community-powered threat data. & \\
\midrule
YARA~\cite{yara2023} & open community & Rules & The pattern matching Swiss knife for malware researchers.& INCONCLUSIVE\\
\midrule
OpenTPX~\cite{OpenTPX2023} &   LookingGlass   & JSON & Machine to machine and human to machine exchange of & YES\\
& cyber solutions &  & any CTI context worth sharing. & \\
\midrule
VERIS~\cite{veris2023} & open & JSON & VERIS is not only rooted in the examination of evidence & YES \\
& community &  & and post-incident analysis, but also designed  & \\
& & & to provide metrics useful to risk management. & \\
 \midrule
OPENIOC~\cite{gibb2013openioc} & Mandiant Inc., & XML & Multiple IoCs can be combined using Boolean logic & NO \\
& open community &  & to define a specific malware sample or family. & \\
 \midrule
TLP~\cite{TLP2023} & NISCC (UK)/ & color  & It is based on the concept of the originator labeling information & YES\\
& US-CERT & coding & with one of the 4 colors to indicate what further dissemination, & \\
& & & if any, can be undertaken by the recipient. & \\
 \bottomrule
\end{tabular}
\label{tab1}
\end{center}
\end{table*}

\subsubsection{Formats}
According to the data formatting, CTIs can be categorized into several classes~\cite{farnham2013tools}. There are 2 groups and organizations behind some of these formats, namely the MILE and MITRE.

The MILE (Managed Incident Lightweight Exchange) Working Group\footnote{\url{https://datatracker.ietf.org/wg/mile/documents/}} worked on the standards for exchanging incident data. The special interest of the group was about defining indicators and incidents under a specific data format. They also worked on the standards for the data-exchange. The MILE group has published a series of standards for CTI which includes:

\begin{itemize}
\item IODEF (Incident Object Description Exchange Format)  \cite{danyliw2007incident} is a CTI data format in which messages are organized in a human-readable way, and not a machine format. IODEF is defined by Request For Comments (RFC) 5070 standard. It is an XML based standard used to share incident information by Computer Security Incident Response Teams (CSIRTs). The IODEF data model uses object-oriented data structure and includes 47 classes used to define incident data. The classes cover a wide range of information including Contact, Monetary Impact, Time, Operating System, Application, etc.
\item IODEF-SCI (IODEF for Structured Cyber Security Information) \cite{takahashi2014incident} is a standard proposed by the MILE working group which extends the IODEF standard to add support for additional data such as attack patterns, platform related info, vulnerabilities, weaknesses, countermeasures, event logs, severeness, etc.
\item RID (Real-time Inter-network Defense) \cite{moriarty2012real} is a standard (defined in RFC 6545) for communicating CTI, especially the IODEF and IODEF-SCI. Transporting RID messages over HTTP/TLS is defined in RFC 6546. RID functions via 5 types of messages: Acknowledgement, Query, Request, Result, and Report.
\end{itemize}

The Mitre Corporation\footnote{https://www.mitre.org/} has developed several complimentary CTI standards that each fills different needs for a CTI management system; CybOX \cite{barnum2012cybox}, STIX \cite{jordan2021stix}, and TAXII \cite{jordan2021taxii}; details of which are provided below:
\begin{itemize}
\item CyBOX (Cyber Observable eXpression) \cite{barnum2012cybox} enables automated CTI sharing including details of measurable events and stateful properties, as well as indicator details called observables. The CybOX objects (e.g. File, HTTP Session, Mutex, Network Connection, Network Flow, and X509 Certificate) can be employed by higher level schemes such as the STIX.
\item STIX standard \cite{jordan2021stix} is for defining CTI information including threat details as well as the context of the threat. The 1.0 version of STIX was released in 2013, the 1.1 version released in 2014, and finally the 2.0 version released in 2017. STIX supports 4 cyber threat use cases: analyzing cyber threats, specifying indicator patterns, managing response activities and sharing CTI. STIX employs XML to define threat related constructs such as campaign, exploit target, incident, indicator, threat actor and TTP. Also, standard definition also includes extensions with other standards such as TLP, OpenIOC, Snort, and YARA.
\item TAXII\cite{jordan2021taxii}  supports sharing of CTI data. 
The first draft of the TAXII specification is released in 2012. TAXII is designed to be flexible, supporting several sharing models including variations of ‘hub and spoke’ as well as ‘peer to peer’, allowing for push or pull transfer of CTI data. The models are supported by 4 main services: Discovery, feed management, inbox, and poll.
\end{itemize}


Others:
\begin{itemize}
\item OTX (Open Threat eXchange) \cite{OTX2023} data format. Here, the format to share information about threats for the OTX community is called ``pulses,'' which provides a summary of the threat, a view into the software targeted, and the related IoC to be used for detecting the threats.
\item YARA (Yet Another Recursive/Ridiculous Acronym) \cite{yara2023} is a tool used to identify and classify malware samples based on custom rules created in the PhishER platform. A rule is a description based on textual or binary patterns. YARA rules are pattern-matching rules used to identify malware. YARA is used by incident responders, threat hunters, and malware forensic analysts, and helps identify and classify malware samples.
\item OpenTPX (Open Threat Partner eXchange)~\cite{OpenTPX2023} is a JSON format to exchange machine-readable CTI as well as network security related information. OpenTPX is based on practical experience building highly scalable threat intelligence analysis and management systems deployed in real-world scenarios. It conveys all aspects of threat intelligence, threat analysis, threat mitigation and network security operation necessary for multiple security and threat intelligence use cases. Specification, data model, data schema and supporting tools are available at.

\item VERIS (Vocabulary for Event Recording and Incident Sharing) \cite{veris2023} was proposed in response to the greatest challenge of the cybersecurity: Lack of quality information. VERIS tackles this challenge by guiding organizations in collecting and sharing the useful CTI information. As such, to provide a common language for describing CTI (structured, repeatable), VERIS is designed by describing a set of metrics. 
\item OpenIOC (Open Indicators of Compromise) \cite{gibb2013openioc} was introduced by Mandiant Inc. in 2011 to be used in their own products, but then has also been released as an open standard. OpenIOC is mainly devised for tactical-CTI and provides definitions for technical details over 500 indicator terms. New terms can be easily added due to separation of the terms for the main schema. The terms are in general host-based including titles of file, driver, disk, system, process, registry, etc. IoC definitions are stored in an XML format.
\item TLP (Traffic Light Protocol) \cite{TLP2023} is a very straight forward and simple protocol which comes from the United States Computer Emergency History (US-CERT, 2013). TLP is used to control what can be done with shared information, based on a coloring scheme. Shared CTI information is tagged with one of the 4 colors white, green, amber or red: White CTI can be distributed without restriction. Green CTI can be shared within the sector (community), but not publicly. Amber CTI may only be shared with members of own organization. RED CTI may not be shared. Due to its` simplicity, TLP can be used via verbally, e-mail, or incorporated into an overall CTI framework.
\end{itemize}

\subsubsection{IoT Implementability}
XML (Extensible Markup Language) is a markup language, designed to store and transport data. On the other hand, JSON (JavaScript Object Notation) is a way of representing objects and used as  a lightweight data-interchange format. In general, JSON is preferred for IoT applications since it can self-describe and is more programmatic, where XML was made for document mark up like HTML \cite{adafruit2023}.

\subsubsection{Summary}
Table~\ref{tab1} summarizes all CTI data formats available today, in a comparative way by classifying according to these categories: Vendor/owner,  data format/protocol, and highlights of each format, and IoT implement-ability. 

\subsection{CTI Exchange Platforms}
Table~\ref{tab2} summarizes all CTI Exchange Platforms available today, in a comparative way by classifying according to these categories: open-source compatibility, latest update year,  and country/region. 


\begin{table}[htbp]
\footnotesize
\caption{Comparison of CTI Exchange Platforms}
\begin{center}
\begin{tabular}{cccc}
\toprule
\textbf{Country/}&\multicolumn{3}{c}{\textbf{Available CTI Exchange Platforms}} \\
\cmidrule{2-4}
\textbf{Region} & \textbf{\textit{Platform name}}& \textbf{\textit{Open-Source}}& \textbf{\textit{Last update}}\\
\midrule
\multirow{7}{*}{Europe} & Abusehelper & \cmark & 2019 \\
 & EclectIQ & \xmark & 2022 \\
 & IntelMQ & \cmark & 2022 \\
 & Megatron & \cmark & 2017 \\
 & MISP & \cmark & 2022 \\
 & N6 & \cmark & 2021  \\
 & Warden & \dmark $^{\mathrm{a}}$ & 2022 \\
\midrule
\multirow{3}{*}{US} & CIF  & \cmark & 2022 \\
 & CRITS & \cmark & 2019 \\
 & Celerium & \xmark & 2022 \\
\bottomrule
\multicolumn{4}{l}{$^{\mathrm{a}}$No clear information exists!}
\end{tabular}
\label{tab2}
\end{center}
\end{table}

After the CTIs are created by the frameworks introduced in Section~\ref{frameworks} according to the formats summarized in Table~\ref{tab1}; they need to be distributed in a seamless way. As such, CTI exchange platforms provided in Table~\ref{tab2} caters this need.










\section{Towards a CTI platform for IoT}\label{CTIforIoT}

\begin{figure}[t]
\centerline{\includegraphics[width=0.49\textwidth]{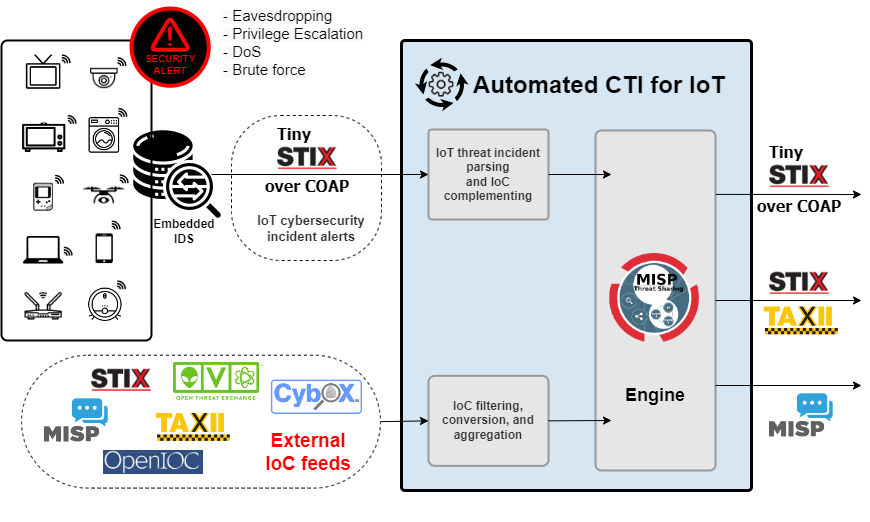}}
\caption{Automated CTI for IoT.}
\label{fig:cti4iot}
\end{figure}

As shown in previous sections, there is a wide variety of CTI standards, formats, and frameworks. However, this heterogeneity creates several challenges related to interoperability, interaction, cooperation, and data sharing and exchange. These challenges result in increased complexity in the management, monitoring, and analysis particularly in IoT systems which usually consist of  a diverse range of devices connected to the Internet, making them vulnerable to numerous threats. On the other hand, all mentioned solutions for CTI management fail to address the potential limitations imposed by the IoT ecosystem, such as resource constraints.
Considering these factors, we propose the architecture of a new CTI management platform  designed specifically for IoT scenarios, as depicted in Figure \ref{fig:cti4iot}. Our proposed system takes into account an IoT environment where low power devices are equipped with intrusion detection capabilities. Therefore, these devices need to generate and process customized lightweight IoCs.
Further details regarding the architecture and technological decisions for the IoT scenario are presented in the upcoming subsections.

\subsection{Scenario}

In an IoT scenario, it is important to establish an effective system capable of collecting IoCs from various sources and in different formats. This system should also be able to convert IoCs into a standardized format and facilitate the sharing with other parties based on specific requirements and formats. This is where a cloud-based CTI engine plays a vital role. CTI engine operating in the cloud can collect IoCs from diverse sources, including security appliances, log files, and threat intelligence feeds, and translate them into a common format that is easily comprehensible and actionable for security personnel.

Furthermore, certain IoT devices can be enabled to generate and distribute lightweight IoCs, primarily through a local Intrusion Detection System (IDS). This capability proves especially valuable for IoT devices situated in remote or inaccessible locations, where they may have limited or intermittent connectivity to a centralized CTI engine.
On the other hand, some IoT devices may be enabled to receive and consume lightweight IoCs for specific purposes, such as updating IDS rules. This empowers these devices to take proactive measures in identifying and mitigating security breaches while effectively responding to potential threats.

\subsection{IoC data model: tinySTIX}

Given that STIX is the predominant and preferred standard for IoC definition, our proposal is to utilize it as a foundation for creating a lightweight version specifically designed for IoT devices, which we call tinySTIX.

Typically, each STIX object occupies hundreds of bytes on disk, which might be considered negligible in traditional networks. However, in the context of an IoT network where multiple IoCs need to be processed and broadcasted daily, this can significantly contribute to the overall power consumption of the network. As the battery level of end-devices represents a critical Key Performance Indicator (KPI) in IoT networks, it becomes imperative to find ways to reduce the size of the STIX messages transmitted.
To minimize the amount of data that IoT devices need to process and transfer, we propose taking action in four key aspects: (i) Controlled vocabularies and property names, (ii) object representation, (iii) security services, and (iv) application protocol. By playing with these aspects, we can effectively decrease the size of the STIX messages transmitted over IoT networks.

\subsubsection{Controlled vocabularies and property names}
Each STIX object consists of a set of properties, comprising both required and optional fields. These properties define various information about the object, such as its type, title, description, and related objects. Certain properties are defined with the requirement of selecting values from predefined lists of keywords. These lists, known as controlled vocabularies, establish a standardized set of terms to describe the represented data. In tinySTIX, controlled vocabularies from STIX are transformed into lists of integer values. Each term within the vocabulary is assigned a unique integer value.
Examples of properties with controlled vocabularies in STIX are: 'type,' 'spec\_version,' 'indicator\_types,' 'relation\_type,' 'pattern\_type,' 'pattern\_version,' 'malware\_types.' The same approach is adopted for property names (keys), where each key is assigned a unique integer value. These integer values are used to replace the corresponding keywords in the IoCs. 

\subsubsection{Object representation}
Unlike STIX v2, which is in JSON format,  tinySTIX uses CBOR (Concise Binary Object Representation) \cite{bormann2013concise} as standard for the object representation. CBOR is preferred to JSON in our scenarios because: (i) CBOR uses binary representation to encode data, making it more compact and efficient to encode and decode; (ii) CBOR has a more compact representation compared to JSON, so it requires less space to store data, and less bandwidth to transmit it, and for this reason CBOR can be decoded faster than JSON, making it more suitable for use in low-powered devices or in real-time communication; (iii)  CBOR provides a way to define custom data types and extensions, making it more flexible than JSON, which only supports a limited set of data types; and (iv) CBOR permits integer key names, while JOSN requires key names to be strings only.

\subsubsection{Security services}
The STIX standard does not itself specify any security protocols for providing digital signatures, encryption, and key management. Since tinySTIX is based on CBOR, it will require the adoption of COSE (CBOR Object Signing and Encryption) \cite{schaad2017cbor} for encryption and signing of CBOR-based objects. COSE provides a range of cryptographic primitives, such as digital signatures and public key encryption, making it well-suited for use in decentralized systems, where trust is established through cryptographic mechanisms. Also, COSE  is designed for use in resource-constrained environments, such as IoT devices and other embedded systems, where processing power and memory are limited.

\subsubsection{Application protocol}
The STIX standard commonly utilizes TAXII as the protocol for exchanging and sharing IoCs. However, given that our proposed tinySTIX relies on CBOR as its representation method, it is more appropriate to employ the CoAP (Constrained Application Protocol) standard for message exchange in IoT scenarios. In fact, CoAP, like CBOR and COSE, is specifically designed for constrained environments. It is a lightweight protocol optimized for devices with limited processing power, memory, and energy resources. 
CoAP primarily operates with the Request/Response (R/R) Model. 
In this model, a CoAP client sends a request to a server. The request can include a method along with one or more options, which provide additional information about the request, such as the URI, content format, or token. These request and response messages are transmitted over UDP or, optionally, over a reliable transport protocol like TCP.

\begin{figure}[th!]
\centerline{\includegraphics[width=0.49\textwidth]{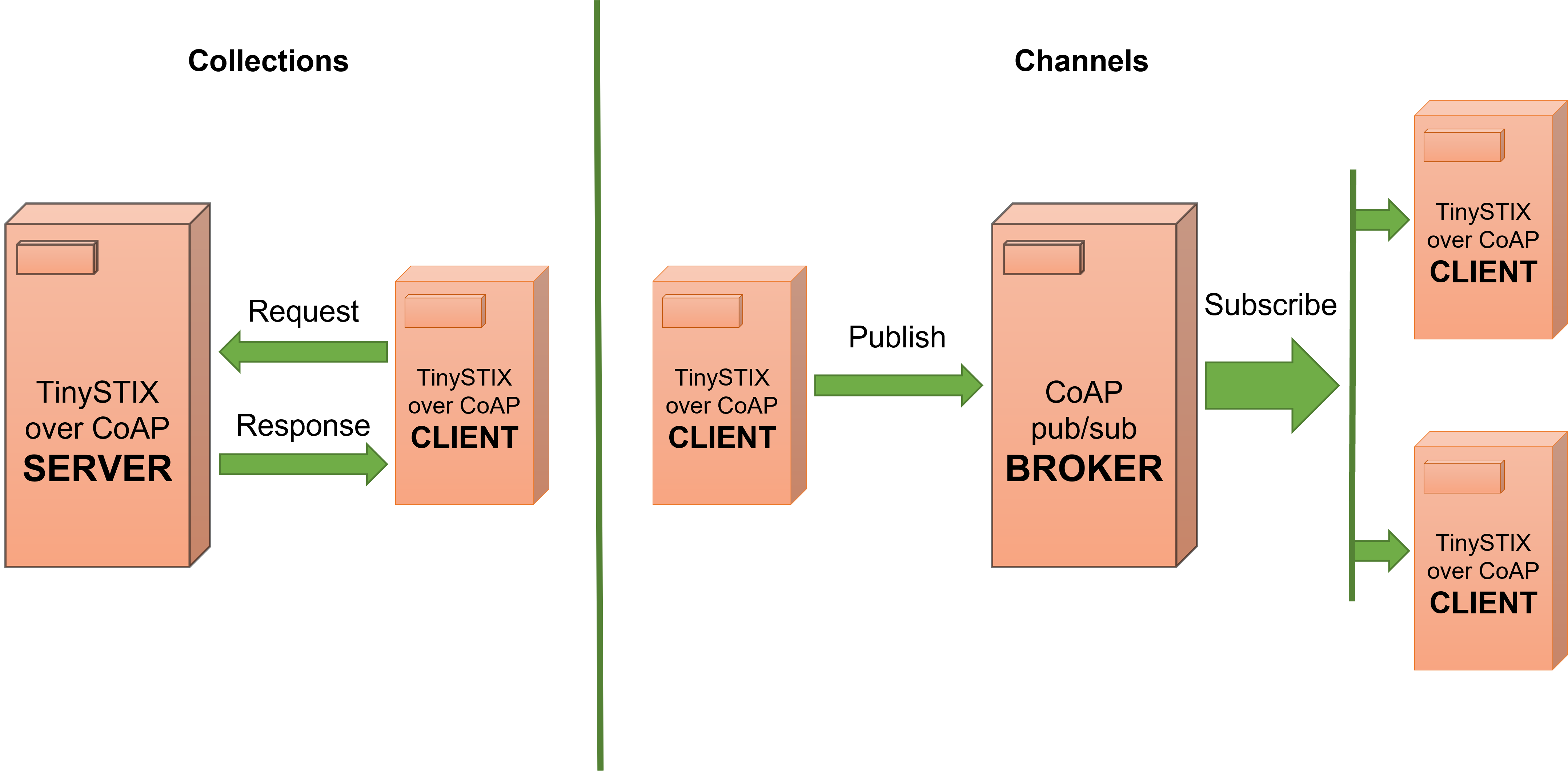}}
\caption{Collections and channels in tinySTIX over CoAP.}
\label{fig3}
\end{figure}

The CoAP R/R Model also supports asynchronous communication through the use of Confirmable and Non-confirmable messages. In addition, a recent CoAP extension supports the Publish-Subscribe Model, which involves a Broker node that enables store-and-forward messaging among multiple nodes. This approach is particularly useful for nodes with limited reachability, as it enables simple and efficient many-to-many communication.
By utilizing the features offered by CoAP, we can easily translate the two CTI sharing models, namely collections and channels provided by TAXII, into corresponding models in CoAP for tinySTIX. Figure \ref{fig3} shows the representation of the two  models for tinySTIX over CoAP.

Together with CoAP, our architecture deploys  OSCORE (Object Security for Constrained RESTful Environments) \cite{selander2019object} to ensure end-to-end security for the sharing of tinySTIX objects among peers within IoT networks. OSCORE is specifically designed to be used alongside CoAP, offering authentication, integrity protection, and confidentiality for CoAP messages. 
OSCORE is purposefully engineered to be simple, lightweight, and well-suited for deployment in environments with limited resources, making it an optimal solution for securing the sharing of lightweight IoCs.

\subsubsection{Analysis of IoC size reduction}

In order to assess the effects of adopting tinySTIX, we conducted an evaluation to measure the reduction in IoC size achieved by implementing integer values for controlled vocabularies and property names, as well as utilizing CBOR object representation. We performed the compression tests on IoCs extracted from two publicly available datasets: (i) STIX 2.1 IoCs sourced from MITRE ATT\&CK, and (ii) MISP IoCs obtained from CIRCL.LU.

\emph{MITRE 
ATT\&CK}\footnote{\url{https://attack.mitre.org/resources/working-with-attack/}} is a publicly accessible database of IoCs provided by The MITRE Corporation. These IoCs are generated based on real-world observations of adversary tactics and techniques. The MITRE Corporation has developed their own ATT\&CK data model, which extends the STIX format by introducing custom STIX objects.
For our evaluation, we utilized the dataset represented in STIX 2.1 JSON collections. This dataset contains IoCs related to attacks in three domains: (i) enterprise (17,672 indicators), (ii) mobile (1,699 indicators), and (iii) Industrial Control Systems (ICS) (1,106 indicators). Table \ref{tab:mitre} provides information on the number of IoC types present in each dataset. Prior to analysis, we processed the datasets by removing all non-STIX native fields.

\emph{CIRCL.LU}\footnote{\url{https://www.circl.lu/doc/misp/feed-osint/}} is a database derived from a publicly accessible IoC feed provided by The Computer Incident Response Center Luxembourg (CIRCL). This database comprises indicators obtained from real-world observations of adversary tactics and techniques. It is regularly updated with the latest indicators and currently contains over 1,500 bundles of IoCs. Table \ref{tab:mitre} presents details regarding the number of IoC types included in the database. As the IoCs in this dataset are provided in MISP format, they were initially converted into STIX 2.1 format, with any non-STIX native fields being removed.

\begin{table}
    \centering
    \caption{Number of IoCs per type and dataset }\label{tab:mitre}
    \begin{tabular}{lcccc}
          \toprule
           \multirow{3}{*}{\textbf{Types}} &\multicolumn{4}{c}{\textbf{Number of IoCs per type}} \\
          \cmidrule{2-5}
           &  \textbf{CIRCL.LU} & \multicolumn{3}{c}{\textbf{MITRE ATT\&CK}}  \\
           \cmidrule{3-5}
           & & \textbf{\textit{Enterprise}} & \textbf{\textit{ICS}} & \textbf{\textit{Mobile}} \\
           \midrule
            artifact	&	112	&	-	&	-	&	-	\\
            attack-pattern	&	5	&	719	&	90	&	175	\\
            autonomous-system	&	2	&	-	&	-	&	-	\\
            campaign	&	2	&	-	&	-	&	-	\\
            course-of-action	&	-	&	284	&	84	&	14	\\
            domain-name	&	14746	&	-	&	-	&	-	\\
            email-addr	&	44	&	-	&	-	&	-	\\
            email-message	&	91	&	-	&	-	&	-	\\
            file	&	4141	&	-	&	-	&	-	\\
            grouping	&	100	&	-	&	-	&	-	\\
            identity	&	1051	&	1	&	1	&	1	\\
            indicator	&	165206	&	-	&	-	&	-	\\
            intrusion-set	&	-	&	140	&	16	&	5	\\
            ipv4-addr	&	3604	&	-	&	-	&	-	\\
            malware	&	-	&	508	&	26	&	93	\\
            marking-definition	&	1041	&	1	&	1	&	1	\\
            mutex	&	15	&	-	&	-	&	-	\\
            network-traffic	&	3604	&	-	&	-	&	-	\\
            note	&	17	&	-	&	-	&	-	\\
            observed-data	&	36373	&	-	&	-	&	-	\\
            process	&	14	&	-	&	-	&	-	\\
            relationship	&	3776	&	15777	&	825	&	1391	\\
            report	&	951	&	-	&	-	&	-	\\
            tool	&	-	&	79	&	1	&	2	\\
            url	&	13697	&	-	&	-	&	-	\\
            user-account	&	3	&	-	&	-	&	-	\\
            vulnerability	&	119	&	-	&	-	&	-	\\
            windows-registry-key	&	83	&	-	&	-	&	-	\\
            x509-certificate	&	1	&	-	&	-	&	-	\\

          \bottomrule
    \end{tabular}
\end{table}

We conducted an analysis to assess the impact of employing the two compression mechanisms, and the corresponding results are presented in Table \ref{tab:reduction}. The total reduction achieved is almost 25\% for the CIRCL.LU dataset and around 50\% for the STIX 2.1 MITRE ATT\&CK dataset. The notable disparity in the reduction percentages between the two datasets primarily stems from variations in the amount of content contained within the IoCs. This discrepancy is particularly evident in fields such as ``description'' or ``attack-pattern.'' Furthermore, the employment of binary listing has a more significant impact compared to CBOR object representation in terms of compression.

\begin{table}
    \centering
    \caption{Average percentage reduction of IoC sizes}\label{tab:reduction}
    \begin{tabular}{lcccc}
          \toprule
           \multirow{3}{*}{\textbf{Types}} &\multicolumn{4}{c}{\textbf{Reduction of IoC sizes per type (\%)}} \\
          \cmidrule{2-5}
           &  \textbf{CIRCL.LU} & \multicolumn{3}{c}{\textbf{MITRE ATT\&CK}}  \\
           \cmidrule{3-5}
           & & \textbf{\textit{Enterprise}} & \textbf{\textit{ICS}} & \textbf{\textit{Mobile}} \\
           \midrule
          Integer value keys  &  21,82 & 43,26 & 42,71 & 46,55 \\
          CBOR representation  & 3,52 & 9,74 & 9,75 & 11,00 \\
          \midrule
          Total reduction &  24,58 & 48,79 & 48,29 & 52,43 \\
          \bottomrule
    \end{tabular}
\end{table}

\subsection{CTI engine: MISP4IoT}
After thorough exploration and analysis of state-of-the-art CTI exchange platforms, we have identified MISP~\cite{wagner2016misp} as the most suitable choice for IoT environments. MISP offers versatility, as it is open-source and can be easily extended and customized for IoT purposes. It is regularly updated and provides a range of well-documented APIs. These characteristics align well with our requirements, making MISP an ideal base engine for our architecture.
MISP already supports STIX and TAXII standards, enabling seamless interoperability with external sources of IoCs. However, to accommodate the new ``tinySTIX'' data format and protocols for constrained devices, MISP needs to be extended. In next subsection, we delve into the methods of implementation for integrating these enhancements into MISP.

\subsubsection{MISP with STIX and TAXII}
MISP provides a Python library called the MISP-STIX converter, which facilitates interactions between the MISP standard and STIX formats.  This converter supports two main features: (i) exporting data collections from MISP to STIX 1.X as well as STIX 2.0/2.1 formats; (ii) 
importing STIX content into a MISP Event. More specifically, MISP can generate STIX content from a given event, including its metadata, and the content of the STIX file can be passed to MISP through the corresponding import function.
To implement the MISP TAXII server, MISP utilizes EclecticIQ's OpenTAXII, which is a Python-based API for TAXII services. Upon launching MISP, a MySQL environment is already active. Once the TAXII server is integrated with MISP, it enables the direct upload of received STIX files into MISP for further processing and analysis.

\subsubsection{MISP with tinySTIX}
The proposed architecture requires an extension for MISP to accommodate the tinySTIX data model and seamlessly incorporate the suite of protocols tailored for constrained devices. This extension should include the implementation of an API for CoAP/OSCORE, supporting CBOR serialization (along with COSE) for the serialization of IoCs, as well as the two IoC exchange models, namely collections and channels.

\section{Conclusion}\label{Conclusion}
In this work, we have been seeking answers to one of the most persistent and critical challenges in the cybersecurity industry/research: ``How to fill in the gap of the lacking quality information on CTI?'' To tackle this issue, we conducted a comprehensive analysis of both historical and contemporary CTI data formats and platforms, providing a comparative overview. In addition, we introduced a novel architecture for a CTI platform specifically designed for IoT networks in which energy consumption and compressed data formats are utmost important. We also proposed a new compressed IoC data model called tinySTIX and demonstrated its effectiveness in significantly reducing the data size required to represent IoCs. Our future works will focus on: (i) tinySTIX implementation on IoT devices, (ii) evaluation of the energy consumption associated with generating and transferring tinySTIX IoCs, and (iii) development of an extension for MISP that enables the processing and conversion of tinySTIX IoCs. 







\section*{Acknowledgment}
This work has been supported by the H2020 project ARCADIAN-IoT (\url{https://www.arcadian-iot.eu/}) [G.A. No. 101020259] . 

\color{black}








\color{black}

\bibliographystyle{IEEEtran}
\bibliography{references}




\end{document}